\documentclass[aip,jcp,superscriptaddress, showpacs,floatfix, nofootinbib,11pt]{revtex4-1}
\usepackage{verbatim}
\usepackage{amsmath}
\usepackage{amssymb,ams fonts,times,natbib}
\usepackage{graphicx}
\usepackage{epstopdf}
\usepackage{xr}
\usepackage{color}

\begin{document}

\title{Probing Polariton Dynamics in Trapped Ions with Phase-Coherent Two-Dimensional Spectroscopy}
\author{Manuel Gessner}
\affiliation{Physikalisches Institut, Albert-Ludwigs-Universit\"at Freiburg, Hermann-Herder-Stra\ss e 3, 79104 Freiburg, Germany}
\author{Frank Schlawin}
\affiliation{Physikalisches Institut, Albert-Ludwigs-Universit\"at Freiburg, Hermann-Herder-Stra\ss e 3, 79104 Freiburg, Germany}
\author{Andreas Buchleitner}
\affiliation{Physikalisches Institut, Albert-Ludwigs-Universit\"at Freiburg, Hermann-Herder-Stra\ss e 3, 79104 Freiburg, Germany}
\affiliation{Freiburg Institute for Advanced Studies, Albert-Ludwigs-Universit\"at Freiburg, Albertstra\ss e 19, 79104 Freiburg, Germany}

\begin{abstract}
We devise a phase-coherent three-pulse protocol to probe the polariton dynamics in a trapped-ion quantum simulation. In contrast to conventional nonlinear signals, the presented scheme does not change the number of excitations in the system, allowing for the investigation of the dynamics within an $N$-excitation manifold. In the particular case of a filling factor one ($N$ excitations in an $N$-ion chain), the proposed interaction induces coherent transitions between a delocalized phonon superfluid and a localized atomic insulator phase. Numerical simulations of a two-ion chain demonstrate that the resulting two-dimensional spectra allow for the unambiguous identification of the distinct phases, and the two-dimensional lineshapes efficiently characterize the relevant decoherence mechanism. 
\end{abstract}

\maketitle

\section{Introduction}
Polaritons - hybrid quasiparticles originating from the strong coupling between light modes and matter excitations - represent an important research area, both for fundamental and practical reasons: Their bosonic character may lead to Bose-Einstein condensation \cite{Deng10, Plumhof14}, and the larger coherence length of the light modes may enhance the carrier mobility in organic semiconductors \cite{Orgiu14}.
While originating primarily in quantum-optical settings such as Josephson junctions and arrays of coupled cavities \cite{Hartmann,QPTL,Bose,Fazio}, polaritonic excitations are now also studied in molecular systems \cite{Schwartz13}.  Alternatively, trapped ions can be used to study the properties of strongly correlated systems under well-controlled conditions and with manageable, slower time scales than solid-state systems \cite{Wineland13,CWReview,RoosReview,Schaetz}.

Polaritonic systems can be modeled in ion trap experiments by coupling the electronic and vibrational degrees of freedom of the ions confined by a harmonic potential. The atomic excitations are encoded into the ions' electronic states, and the local vibrational modes of the ions, described by quantum mechanical harmonic oscillators, take on the role of the light modes. The Coulomb repulsion between the ions leads to couplings of the local phonons. Finally, a suitably chosen external laser field realizes Jaynes-Cummings type interactions\cite{Jaynes63,Leibfried} between each ion's electronic state and its local phonon mode \cite{Ivanov09, Toyoda13}. The full system is then described by a Jaynes-Cummings-Hubbard model \cite{Ivanov09}, in which the total number of electronic and vibrational excitations is conserved.

For filling factor one ($N$ excitations on $N$ sites), the ground state varies, with the detuning of the external laser field, from a phonon superfluid phase, in which vibrational excitations are delocalized over the entire chain, to an atomic Mott insulator phase, in which electronic excitations reside on individual ions. While all excitation energy is stored in the vibrational degree of freedom, in the former case, it is fully absorbed by the electronic degree of freedom, in the latter. In the intermediate regime, the system exhibits a polaritonic phase of coupled atom-phonon excitations. When the filling factor is varied, the model allows for further phases, such as a polariton glass \cite{Fazio}.

In this work, we devise an interaction scheme which induces coherent transitions between these two phases. This interaction will allow us to design pulse sequences \cite{Gessner14,Schlawin14} for multidimensional nonlinear spectroscopy of the polariton dynamics. In contrast to conventional spectroscopic techniques, our interaction does not change the number of excitations in the system, but rather switches between electronic or phononic character of the excitations. In a phase-coherent three-pulse sequence, the system evolves through both the superfluid and the insulator quantum phases, as well as a coherent superposition of both. We will see that nonlinear spectroscopy is particularly well suited to identify the relevant excitation processes, as well as the dominant decoherence mechanisms via the observed two-dimensional lineshapes. To ease scalability towards large system sizes, in which quantum phase transitions become most pronounced, we renounce of single-ion addressability in our spectroscopic protocol, which, however, could be exploited to add spatial resolution to the obtained nonlinear spectra \cite{Gessner14,Schlawin14}.

\section{Background}
\subsection{Dynamics and laser control of trapped ions}\label{sec.introions}
The dynamics of polaritonic systems can be simulated by subjecting a collection of trapped ions to appropriate laser fields \cite{Ivanov09,Toyoda13}, taking into account both the electronic and motional states of all the ions. As mentioned before, the relevant electronic states of the trapped ions are modeled as non-interacting spin-$1/2$ systems \cite{Haeffner}, while the motion is determined by their common trap potential, which, to a good approximation, is described by a three-dimensional harmonic oscillator \cite{Leibfried}. The analogy to polaritonic systems is achieved by identifying the phonons in the ion trap with the photons of coupled cavities. In the remainder of this article we refer to the actual physical degree of freedom, i.e., the phonons, rather than the mathematically equivalent photons in the case of an optical polaritonic system.

Near their motional ground state, the interplay between the potential and the strong Coulomb repulsion of the ions determines the chain geometry. The equilibrium positions of the ions are determined by the ratio of the trap frequencies $(\nu_x,\nu_y,\nu_z)$ along the different spatial directions. A linear chain arrangement is achieved when, e.g. $\nu_x\approx\nu_y\gg\nu_z$, and allows for convenient laser addressing of individual ions, with typical inter-ion spacings of $5-15\,\mu m$ and typical trap frequencies on the order of $\nu_x,\nu_y\approx 1-10\,$MHz and $\nu_z\approx 0.1-1\,$MHz. In the following, we will consider such a linear chain of laser-cooled ions, which allows us to assume that the ions' displacement from their equilibrium positions is much smaller than their separation. Employing a second-order expansion\cite{James} in terms of the displacement along the $x$-direction leads to the Hamiltonian (we set $\hbar = 1$)
\begin{align}\label{eq.H0}
H_0 &= \omega \sum_{k = 1}^N \sigma_k^+ \sigma_k^- + \sum_{k = 1}^N (\nu_x + \omega_k) a_k^{\dagger} a_k  + \sum_{\substack{k,l=1\\k < l}}^N t_{kl} \left( a^{\dagger}_k a_l +  a^{\dagger}_l a_k \right), 
\end{align}
where $a_k^{\dagger}$ creates a local phonon along the $x$-direction at the ion $k$, and $\sigma^{\pm}_k$ is the spin-$1/2$ ladder operator at site $k$. The electronic states of all the ions are described by the same transition frequency $\omega$.
Due to large spatial separation between adjacent ions, direct transfer of electronic excitations between the ions may be neglected. However, the Coulomb repulsion of the ions affects the motional degrees of freedom in Eq. (\ref{eq.H0}): It is responsible for both, modulations of the local oscillator frequencies $\omega_k=-\sum_{j\neq k}t_{kj}$ and the phonon-phonon couplings $t_{kl} = \nu_x\beta/(2|u_k-u_l|^3)$, where $\beta=\nu_z^2/\nu_x^2$ is the ratio between the confining trap frequencies in axial $(z)$ and radial $(x)$ direction, and $u_l$ the ions' equilibrium positions in units of $l_0 = (e^2 / (m \nu_z^2))^{1/3}$. \cite{James}

An essential tool for the control of the quantum state of trapped ions are coherent laser interactions \cite{Leibfried,Haeffner,Schindler}. The interaction Hamiltonian, generated by a continuous-wave laser, may be written as
\begin{align}\label{eq.H_int}
H_{\text{int}} &= \frac{\Omega}{2} \left( \sigma_k^+ + \sigma_k^- \right) \left( e^{i (\vec{k}_L \vec{r}_k -  \omega_L t)} + e^{- i (\vec{k}_L \vec{r}_k - \omega_L t)} \right), 
\end{align}
and describes the interaction of the $k$-th ion (at the position $\vec{r}_k$) with a laser beam of frequency $\omega_L$, wavevector $\vec{k}_L$, and Rabi frequency $\Omega$. For laser-cooled ions of low kinetic energy, the trap potential is well approximated by a static harmonic potential \cite{Leibfried}, and we may express the ion position in terms of the harmonic ladder operators. For example, when $\vec{k}_L$ coincides with one of the motional axes, e.g., the $x$-direction, in the case of a single ion, we obtain $\vec{k}_L \vec{r}_k = |\vec{k}_L| x_k=\eta(a_k + a^{\dagger}_k)$, with the bosonic ladder operators as defined above, and the Lamb-Dicke parameter $\eta=k/\sqrt{2m\nu_x}$, which depends on the frequency $\nu_x$ of the harmonic motion. In general, for arbitrary numbers of ions, tuning the frequency $\omega_L$ allows to engineer effective couplings between the electronic and vibrational degrees of freedom, as well as between the electronic states of different ions \cite{Haeffner,PorrasCiracSpins}.
To see this, it is important to notice that the electronic spectrum of the ions is modulated by vibrational sidebands [as can be seen in Eq.~(\ref{eq.H0})], which originate in the ions' motion around their equilibrium positions. In analogy to Stokes and anti-Stokes lines in Raman spectroscopy \cite{CT}, these resonances are found when the laser detuning coincides with multiples of the frequency of the harmonic motion.
 
Using tunable narrow-band lasers, specific sidebands can be resonantly selected. This effectively creates coupled electronic and vibrational excitations: Driving the first red (blue) sideband leads to laser-induced transitions which create an electronic excitation while a phonon is removed (created). Hence, the dynamics of, e.g., the red sideband is described by a Jaynes-Cummings Hamiltonian which induces couplings of the type $\sim \sigma_k^+ a_k+\sigma_k a_k^{\dagger}$. A more detailed treatment that includes the contribution of various harmonic modes is described in appendix~\ref{appendixA}.

\subsection{The polaritonic Hamiltonian}
In this paper, we consider the scenario considered in ref. \cite{Ivanov09}. To couple motional and electronic degrees of freedom, and thereby create polaritonic excitations, one irradiates the entire ion chain with a travelling-wave laser, oriented along the radial $x$-direction, whose frequency is chosen as $\omega_L=\omega-\nu_x+\Delta$, i.e., the laser is detuned from the radial red sideband transition by $\Delta$. \cite{Toyoda13}

In the Lamb-Dicke limit (c.f. appendix~\ref{appendixA}), and after performing the rotating wave approximation, the dynamics of $N$~ions is effectively described in an appropriate rotating frame by the Hamiltonian \cite{Ivanov09}
\begin{align}
H = \sum_{k=1}^N \omega_k a_k^{\dagger} a_k + \sum_{\substack{k,l=1\\k < l}}^N t_{kl} \left( a^{\dagger}_k a_l +  a^{\dagger}_l a_k \right) + \Delta \sum_{k=1}^N \sigma_k^{+} \sigma_k^- + g \sum_{k=1}^N \left( \sigma_k^+ a_k + \sigma_k^- a^{\dagger}_k \right), \label{eq.H}
\end{align}
where, as before, $a_k^{\dagger}$ creates a local phonon along the $x$-direction at site $k$, and $\sigma^{\pm}_k$ is the spin-$1/2$ ladder operator of ion $k$. The coupling between spin and phonon degrees of freedom is denoted with an effective Rabi frequency $g$. The detuning $\Delta$ from the red sideband transition determines the effective local spin's energy splitting. The validity of the above Hamiltonian is limited to the case where all parameters $\beta$, $\Delta$, and $g$ are small \cite{Ivanov09}, for the following reasons: The ratio of squared trap frequencies $\beta$ must be small to justify the quadratic expansion of the Coulomb interaction term which lead to the motional Hamiltonian~(\ref{eq.H0}). The detuning $\Delta$ must be significantly smaller than the trap frequency $\nu_x$ to ensure the validity of the rotating wave approximation which was employed to obtain Eq.~(\ref{eq.H}). Finally, when the laser-induced atom-phonon coupling becomes too large, multiple phonons can be created or destroyed in one electronic transition, thus higher-order terms would have to be included in the interaction term in Eq.~(\ref{eq.H}).

\begin{figure}
\includegraphics[width=0.99\textwidth]{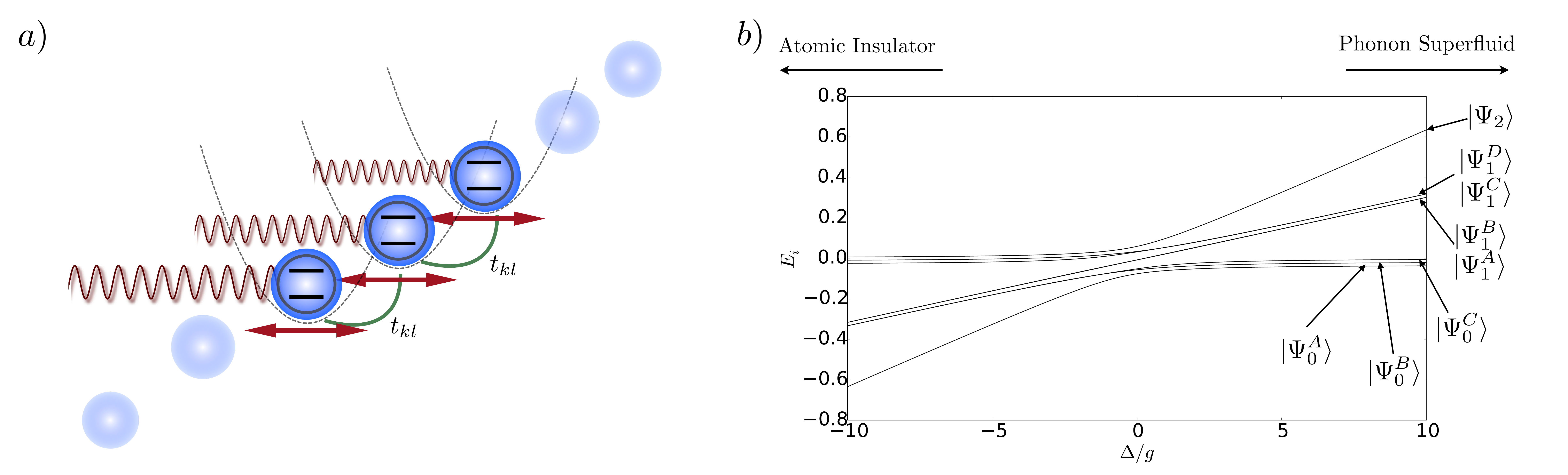}
\caption{(a) Schematic representation of the trapped-ion quantum simulation of polaritonic systems. Laser fields couple the electronic states of the ions to their motional degrees fo freedom, which in turn are coupled by the common trap potential. (b) Two-excitation manifold of the Hamiltonian~(\ref{eq.H}) for $N=2$ ions as a function of the ratio between detuning and interaction strength $\Delta / g$. The eigenstates are indicated, their subscripts denote the number of spin excitations, and the superscripts label the individual states within a spin excitation submanifold.
}
\label{fig.levelscheme}
\end{figure}

We reiterate that the coupling between spins and phonons is of polaritonic (rather than polaronic) nature, since it leads to an exchange of atomic and phononic excitations: One easily verifies that the total number of excitations $N = \sum_k ( a^{\dagger}_k a_k + \sigma^+_k \sigma^-_k )$ is a constant of motion of the Hamiltonian~(\ref{eq.H}). \cite{Ivanov09} Due to the interaction between phononic and electronic degrees of freedom, an initial spin excitation localized on a specific ion will be transferred to the phononic subspace, spread due to the Coulomb interaction, and transfer back to the electronic subspace. Our present analysis aims to understand the dynamics in a given $N$-excitation manifold of the system. In particular, we restrict to the case where the number of excitations equals the number of ions, i.e., for a filling factor one.

In this case, the ground state of the system's $N$-excitation manifold 
for $-\Delta/g\gg 1$ is given by the uncorrelated \textit{atomic insulator} (atI) state 
\begin{align}
|\mathrm{atI}\rangle = |\uparrow\rangle^{\otimes N}\otimes |0\rangle^{\otimes N}
\end{align}
with localized spin excitations. Here, $|\uparrow \rangle$ denotes the excited electronic state, and $|0\rangle$ the phononic vacuum. 

Conversely, for $\Delta/g\gg 1$, the lowest-energy state is characterized by excitations of the motional eigenmode of lowest energy \cite{James}, created by the operator $b^{\dagger}_1=\sum_kc_{k1}^* a^{\dagger}_k$. This state is denoted the \textit{phonon superfluid} (phSF) state,
\begin{align}
|\mathrm{phSF}\rangle=|\downarrow\rangle^{\otimes N}\otimes \frac{1}{\sqrt{N!}}\left(b^{\dagger}_1\right)^N|0\rangle^{\otimes N}, \label{eq.psi_phSF}
\end{align}
since the excitations are spatially delocalized within the phonon degrees of freedom. In the intermediate regime, $\Delta/g\approx 0$, the system is described by a polaritonic superfluid phase \cite{Irish}. This phase is characterized by a delocalized ground state of polaritonic character, i.e., containing both atomic and phononic excitations \cite{Irish}.

Let us now also briefly discuss the excited states for the simplest nontrivial example: the two-excitation subspace in a two-ion chain. The structure of the eight-dimensional spectral manifold is depicted in Fig.~\ref{fig.levelscheme}b), as a function of the ratio $\Delta/g$. 

If $\Delta/g \gg 1$, the spin Hamiltonian $ \Delta \sum_{k=1}^N \sigma_k^{+} \sigma_k^-$ dominates over all other terms in the Hamiltonian. In the case of the two-ion chain considered here, this Hamiltonian has three eigenvalues $0$, $\Delta$, and $2 \Delta$ (which are determined by the number of excited spins), and consequently we identify three submanifolds of states, which differ in the number of spin excitations: The submanifold $\{|\Psi_0^A\rangle,|\Psi_0^B\rangle,|\Psi_0^C\rangle\}$, 
with the ground state $|\Psi_0^A\rangle$, encompasses all three states with close to zero spin excitations. The first excited-state submanifold contains four states $\{|\Psi_1^A\rangle,|\Psi_1^B\rangle,|\Psi_1^C\rangle,|\Psi_1^D\rangle\}$ with a single spin excitation. Finally, both spins are excited in the highest excited state $|\Psi_2\rangle$. Since our analysis restricts to the two-excitation subspace, the respective spin excitations are complemented by phonon excitations among the two eigenmodes to reach a total of two excitations. 

In the other limit, when $-\Delta/g\gg1$, the same three submanifolds can again be identified, however, their energies are reversed: The state $\vert \Psi_2 \rangle$ now constitutes the ground state manifold, and the zero-spin submanifold $\{|\Psi_0^A\rangle,|\Psi_0^B\rangle,|\Psi_0^C\rangle\}$ yields the highest-excited states. 

In between these two limits, when $\vert \Delta / g \vert \leq 1$, the eigenstates lose their clear character, and in the thermodynamic limit, one expects an abundance of avoided crossings which accompany the quantum phase transition and point to quantum chaotic behavior \cite{Schlawin14}.



\subsection{Phase-coherent two-dimensional spectroscopy}
Nonlinear spectroscopy provides powerful tools to probe the dynamics and decay processes of complex interacting quantum systems \cite{Mukamel}. A general theoretical framework in the context of trapped-ion systems was recently developed \cite{Gessner14, Schlawin14}, which allows for the identification of a variety of phenomena ranging from decoherence-free subspaces \cite{Gessner14} to phonon currents in a nonequilibrium steady-state \cite{Schlawin14} and structural phase transitions in ion crystals \cite{Plenio15}. The two essential ingredients are sequences of short, perturbative pulses separated by tunable time-delays, and a mechanism that allows to post-select the phase signature of particular excitation sequences. In large ensembles, the latter can be achieved via phase matching, which is not possible for trapped-ion systems, due to the limited amount of scattered photons per incoming probing pulse \cite{Gessner14}. A feasible alternative is presented by phase-cycling which was developed to probe nuclear magnetic resonance systems with nonlinear spectroscopy \cite{nmr}. 

In  the following, we shall employ these ingredients to investigate the dynamics induced by Eq.~(\ref{eq.H}).

\section{Results}
\subsection{Probing the $N$-excitation subspace with a phase-coherent three-pulse scheme}
For the present simulation study, we consider the interaction with a global laser field, which illuminates all the ions in the trap with equal intensity and is tuned resonantly to the motional sideband of the collective mode $b_1$. This allows to realize an interaction of the form\cite{Wineland1,Wineland2,Retzger,Roos} (see appendix~\ref{appendixA})
\begin{align}
V_{I} &=i\frac{\widetilde{\Omega}}{2}\left(e^{i\phi}J_+b_1 - e^{-i\phi}J_-b^{\dagger}_1\right) \label{eq.H_I}\\ 
&=i \left( V_I (\phi) - V_I^{\dagger} ( \phi) \right), \label{eq.V_I}
\end{align}
where $J_{\pm}= \sum_{k = 1}^N c_{k 1} \sigma_k^{\pm}$ creates collective excitations among the spins, where the coefficients $c_{k1}$ decompose the collective mode $b_1$ into local contributions as $b_1=\sum_kc_{k1}a_k$. It is possible to switch between the atomic insulator and the phonon superfluid states by inducing $N$ interaction events of the above form, when $b_1$ describes the vibrational eigenmode of lowest energy. These interactions conserve the number of total excitations in the system: Each spin excitation is accompanied by a phonon annihilation, and vice versa. Phase cycling with respect to the phase $\phi$ can distinguish between the different contributions to the total signal. This technique further renders weak, perturbative pulses sufficient to obtain multidimensional spectra combined with a precise time resolution, and allows to post-select the evolution of specific coherences.

\begin{figure}
\includegraphics[width=0.4\textwidth]{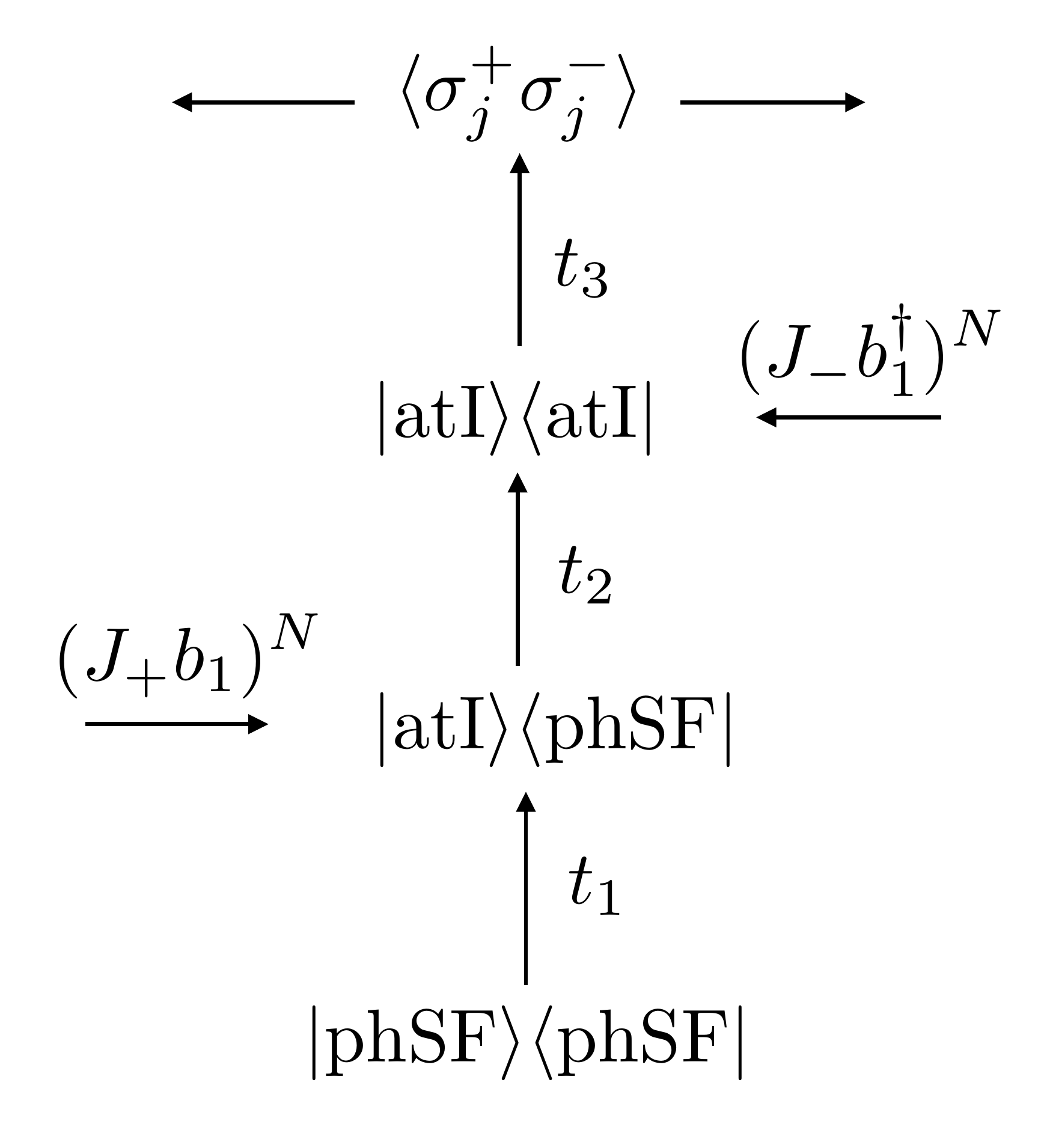}
\caption{Pulse sequence for the nonlinear measurement protocol: The system is prepared in the superfluid phase $|\mathrm{phSF}\rangle$, and evolves freely during $t_1$. Two excitation pulses then transfer the chain into the atomic insulator phase $|\mathrm{atI}\rangle$. Post-selecting the phase $N (\phi_2 - \phi_2)$ allows to separate the excitation process into two steps, and to study the evolution of the coherence $|\mathrm{atI}\rangle \langle\mathrm{phSF}|$ during $t_2$. The readout signal is created at the end of the sequence by fluorescence detection at ion $j$.}
\label{fig.diagram}
\end{figure}

Based on an extension of the single quantum coherence signal \cite{Gessner14}, we design the following excitation scheme (see Fig.~\ref{fig.diagram}):
\begin{enumerate}
\item The system is initialized in the phonon superfluid state $|\mathrm{phSF}\rangle$.
\item During the first time interval, $t_1$, the system evolves in the phonon superfluid state~(\ref{eq.psi_phSF}). This evolution is described by the Liouville space superoperator $\mathcal{G} (t) = \exp [ \mathcal{L} t]$, where the Liouvilian $\mathcal{L}$ contains the coherent evolution according to the Hamiltonian~(\ref{eq.H}), as well as the influence of possible noise sources. Some of the dominant noise sources of trapped-ion experiments are mentioned in the course of our simulations in the next section. We obtain $\mathcal{G} (t_1)|\mathrm{phSF}\rangle\langle\mathrm{phSF}|$.
\item The first excitation pulse populates higher-excited states of the system.
By phase cycling with respect to the phase signature $N (\phi_1 - \phi_2)$, where $\phi_i$ denotes the phase of the $i$-th pulse [see Eq.~(\ref{eq.H_I})], it is possible to post-select those processes, in which the excitation pulse transfers the system into a superposition state of phonon superfluid and atomic insulator. Thus, the contributing terms are described by $(\mathcal{V}_I^{(L)})^N\mathcal{G} (t_1)|\mathrm{phSF}\rangle\langle\mathrm{phSF}|$, where $\mathcal{V}_I^{(L)}\rho=J_+b_1\rho$ represents the Liouville space superoperators associated with an excitation acting only on the left (L) side of the density matrix. Similarly, we introduce $\mathcal{V}_I^{(R)}\rho=\rho J_-b^{\dagger}_1$ The decay of this superposition can be observed during the second time interval $t_2$, and is described by $\mathcal{G} (t_2)(\mathcal{V}_I^{(L)})^N\mathcal{G} (t_1)|\mathrm{phSF}\rangle\langle\mathrm{phSF}|$
\item The phase cycling procedure described above post-selects pathways, in which the second excitation transfers the system fully into a population of the atomic insulator state, whose dynamics is monitored during $t_3$. We obtain $\mathcal{G} (t_3)(\mathcal{V}_I^{(R)})^N\mathcal{G} (t_2)(\mathcal{V}_I^{(L)})^N\mathcal{G} (t_1)|\mathrm{phSF}\rangle\langle\mathrm{phSF}|$
\item Finally, the signal is obtained by measuring the spin population of one of the ions (characterized by the ion's index $j$) via fluorescence \cite{Leibfried,Haeffner,Schindler}.
\end{enumerate}
The measured signal is given by the nonlinear correlation function\cite{Schlawin14}
\begin{align}\label{eq.signal-definition}
S (t_1, t_2, t_3; j) &= \mathrm{tr} \left\{ \sigma^+_j \sigma^-_j \mathcal{G} (t_3) ( \mathcal{V}_I^{(R)} )^N \mathcal{G} (t_2) (\mathcal{V}_I^{(L)})^N \mathcal{G} (t_1) \rho_0 \right\}, 
\end{align}
where $\rho_0 = |\mathrm{phSF}\rangle \langle \mathrm{phSF}|$ denotes the initially prepared superfluid state and $\mathrm{tr}$ the trace. To efficiently analyze this multidimensional signal, we will Fourier transform two of the time delays,
\begin{align}\label{eq.S^13}
S^{(1,3)} (\Omega_1, t_2, \Omega_3; j) &= \int dt_1 \int dt_3 \; e^{i (\Omega_1 t_1 + \Omega_3 t_3)} S (t_1, t_2, t_3; j), 
\end{align}
or
\begin{align}\label{eq.s23}
S^{(2,3)} (t_1, \Omega_2, \Omega_3; j) &= \int dt_2 \int dt_3 \; e^{i (\Omega_2 t_2 + \Omega_3 t_3)} S (t_1, t_2, t_3; j).
\end{align}
Let us discuss the conditions and implications of each of the above steps in further detail. The initialization of the phonon superfluid state can be achieved via a combination of ground-state laser cooling, optical pumping, and well-defined pulses on blue sideband transitions \cite{Leibfried,Gardiner1997}. If the system parameters are chosen such that the initial state $|\mathrm{phSF}\rangle$ actually represents an eigenstate, there is no time evolution during $t_1$. The presence of small contributions of other excited states will become visible in the resulting two-dimensional spectrum. Such contributions would also open up the possibility of removing spin excitations with the first pulse. For $|\Delta/g|\gg1$, the state $|\mathrm{phSF}\rangle$ is very close to an eigenstate of the system: When $\Delta > 0$, it constitutes the ground state, when $\Delta < 0$ the highest excited state, with small corrections due to the finite spin-phonon coupling $g$. In this case, we can neglect contributions from other diagrams, in which, for instance, an initial spin excitation creates additional pathways. These need to be taken into account only when $|\Delta/g|\leq1$, which shall not be our concern in this paper.
Finally, the measurement of spin populations is easily realized by fluorescence detection \cite{Leibfried,Haeffner,Schindler}. Since the spatial decomposition of the addressed eigenmode is contained in the spin excitation operator $J$, the signal can be dependent on $j$, i.e., we obtain spatial information about the relevant vibrational mode only with local readout, without resorting to local excitations. Single-site addressing is in principle possible and is able to enhance the information content compared to conventional nonlinear spectroscopic signals \cite{Gessner14,Schlawin14}. Whereas the spatially resolved readout of the fluorescence light is routinely realized via high-resolution cameras, even for large $N$, the addressing of individual ions with focussed lasers becomes technically demanding for increasing system sizes \cite{Schindler}. Since the macroscopic properties of the quantum phases and the corresponding phase transition is most interesting in this large-$N$ limit, we avoid the usage of focussed excitations to ease scalability.

In the present analysis, we have considered the impulsive limit, which implies that the length of the applied pulses is much shorter than the characteristic time scale of the relevant system dynamics. The dynamics of the effective polaritonic Hamiltonian~(\ref{eq.H}) is determined by the motional frequencies, as well as by the detuning of the optical field. Experimentally this typically leads to relevant time scales on the order of hundreds of microseconds \cite{Toyoda13}. Moreover, the evolution can be slowed down artificially by adiabatically changing the ratio of trap frequencies. These time scales allow for the applications of pulses that, on the one hand, are short compared to the system dynamics, and, on the other hand, are long enough to spectrally resolve individual transitions.

Whereas often the information contained in single quantum coherence signals is also available in technically less demanding Ramsey interferometric measurement sequences \cite{Wineland13}, this is not the case for the present simulation study. In a weak-field approximation, phase-cycling enables us to isolate those events where exactly $N$ excitations are exchanged with the interaction pulses -- of which no clear analogue exists in the context of a Ramsey experiment.

\subsection{$S^{(2,3)}$: Coherence decay of distinct quantum phases and evolution of the atomic insulator state}
\begin{figure}
\includegraphics[width=\textwidth]{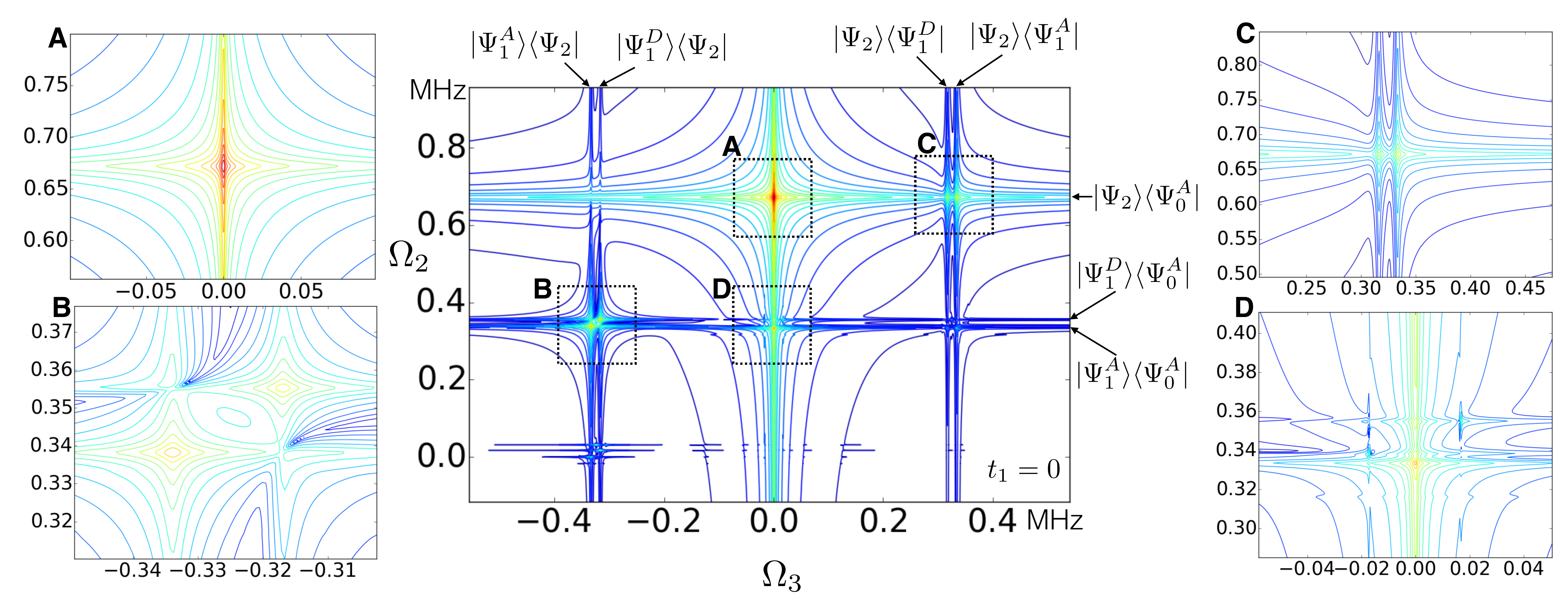}
\caption{Center: Absolute value of the two-dimensional spectrum $|S^{(2, 3)} (t_1 = 0, \Omega_2,\Omega_3)|$ 
(where we omit the site index $j$ in (\ref{eq.s23}), since the fluorescence signal is independent of $j$ for $N=2$) of the 
pulse sequence~(\ref{eq.signal-definition}) in a two-ion chain implementing 
the Hamiltonian~(\ref{eq.H}),
with $\nu_x=1\,$MHz, $\beta=5\,$kHz, $\Delta = 50\,$kHz, $g = 5\,$kHz. The panels A--D zoom into specific resonance structures to highlight the two-dimensional lineshapes.}
\label{fig.f2f3}
\end{figure}

We now present simulations of the three-pulse sequence~(\ref{eq.signal-definition}) with the level scheme in Fig.~\ref{fig.levelscheme} for the ratio $\Delta / g = 10$ at $N=2$. The initially prepared state $|\mathrm{phSF}\rangle$ has strong overlap with the manifold ground state $|\Psi_0^A\rangle$. Small corrections from the first single spin excitation submanifold become visible through oscillations during $t_1$. After two interactions with the first pulse, the strongest contribution to the generated state $|\mathrm{atI}\rangle$ is given by the highest excited state $|\Psi_2\rangle$. In the subsequent time interval $t_2$, the decay of the coherence $|\mathrm{atI}\rangle\langle\mathrm{phSF}|$ can be observed in the two-dimensional spectrum. 

Fig.~\ref{fig.f2f3} displays the two-dimensional spectrum $S^{(2,3)} (t_1 = 0, \Omega_2, \Omega_3)$ (where we omit the site index 
$j$ in (\ref{eq.signal-definition}-\ref{eq.s23}), since the fluorescence signal is independent of $j$ for $N=2$). Our simulations were carried out using the \texttt{qutip} package \cite{qutip}. The parameters used in our simulation are $\nu_x=1\,$MHz, $\beta=5\,$kHz, $\Delta = 50\,$kHz, $g = 5\,$kHz and reflect those of a recently reported experiment \cite{Toyoda13}. To obtain realistic results, we further include the effect of the dominant error source -- collective dephasing of the qubits due to fluctuations in the magnetic field that determines the common electronic resonance frequency \cite{Schindler,Ben,Edo}, described by the dephasing operator $\sigma_z^{\otimes N}$ with decay constant $\gamma = 0.5\,$kHz. Another relevant noise source is caused by fluctuations of the laser intensity of the driving fields\cite{Schindler}, which we do not consider here. In the following, we analyze the position and two-dimensional lineshapes of cross-peaks, which are direct traces of the character (localized or delocalized) of the involved states.

We identify four strong resonances in Fig.~\ref{fig.f2f3}, which are labelled A,B,C,D. All peaks can be uniquely attributed to a specific excitation pathway by extracting the corresponding resonance frequencies along $\Omega_2$ and $\Omega_3$. Resonances A and C at $\Omega_2 \approx 672\,$kHz originate from the coherence $|\Psi_2\rangle \langle \Psi_0^A |$ during the $t_2$-evolution. Hence, it is those two peaks that trace the evolution of the ``macroscopic" coherent superposition of the two distinct phases $(|\mathrm{phSF}\rangle+|\mathrm{atI}\rangle)/\sqrt{2}$. The stronger
resonance A at $\Omega_3 = 0$
signals the evolution of the population $|\Psi_2\rangle \langle \Psi_2 |$ during $t_3$. The states $|\Psi_1^A \rangle$ and $|\Psi_1^D\rangle$ are populated during $t_3$ due to the weak residual coupling of the spin-phonon interaction. Two less pronounced resonances at $\Omega_3 \approx 350$~kHz reveal the coherent evolution of 
superposition states that involve the coherences $|\Psi_2\rangle\langle\Psi_1^{A/D}|$.

Furthermore, the two-dimensional lineshapes of the resonances can be used as a tool to identify the nature of the underlying states. The ability to analyze the two-dimensional lineshapes\cite{HammZanni} represents another distinct advantage of multi-dimensional spectroscopy, and becomes particularly useful in the present context. The influence of the collective dephasing process increases quadratically with the difference of occupation numbers that contribute to the superposition \cite{Palma,Breuer,Fischer}, i.e., we have $\mathcal{G} (t) |\mathrm{atI}\rangle\langle\mathrm{phSF}| \sim e^{- N^2 \gamma t}$, and therefore is most pronounced for the decay of electronic coherences between the ground state manifold and the highest excited state. When both states contain exactly the same number of spin excitations, they evolve in a decoherence-free subspace \cite{DFS}. Here, however, the non-zero coupling between spins and phonons perturbs the symmetry of the subspace and will always lead to finite linewidths. Hence, both resonances A and C are broadened strongly along $\Omega_2$ (where the discrepancy of spin excitations is two), and much less along $\Omega_3$ (where the states differ by no more than a single spin excitation). As discussed before, the resonance A 
represents the population $|\Psi_2 \rangle \langle \Psi_2 |$ during $t_3$, i.e., it evolves in a decoherence-free subspace. In contrast, the resonance C is also broadened along $\Omega_3$, but due to the smaller difference in the spin population between $|\Psi_1^{A/D} \rangle$ and $|\Psi_2 \rangle$, the broadening is less pronounced compared to the 
strong broadening along $\Omega_2$.


Resonances B and D at $\Omega_2 \approx 350$~kHz pertain to processes in which the two interactions with the first pulse only create a single spin excitation -- again, due to the residual spin-phonon coupling -- and the 
system therefore evolves in a coherence between one of the two single-spin states $|\Psi_1^A \rangle$ and $|\Psi_1^D\rangle$ and the initially prepared ground state $|\Psi_0^A\rangle$. 
At resonance B, the second pulse then creates two spin excitations on the \textit{bra} side of the density matrix, such that the 
system evolves in the coherence $|\Psi_1^{A/D} \rangle \langle \Psi_2 |$ during $t_3$. Consequently, the difference of spin excitations 
remains unchanged during both propagation 
intervals, and the resonances are broadened equally along both $\Omega_2$ and $\Omega_3$. Finally, the resonances in the close-up D  
represent a population during $t_3$, such that they are only broadened along $\Omega_2$ -- just like resonance A. 

\subsection{$S^{(1,3)}$: Time evolution of populations of the two quantum phases}

\begin{figure}
\includegraphics[width=.55\textwidth]{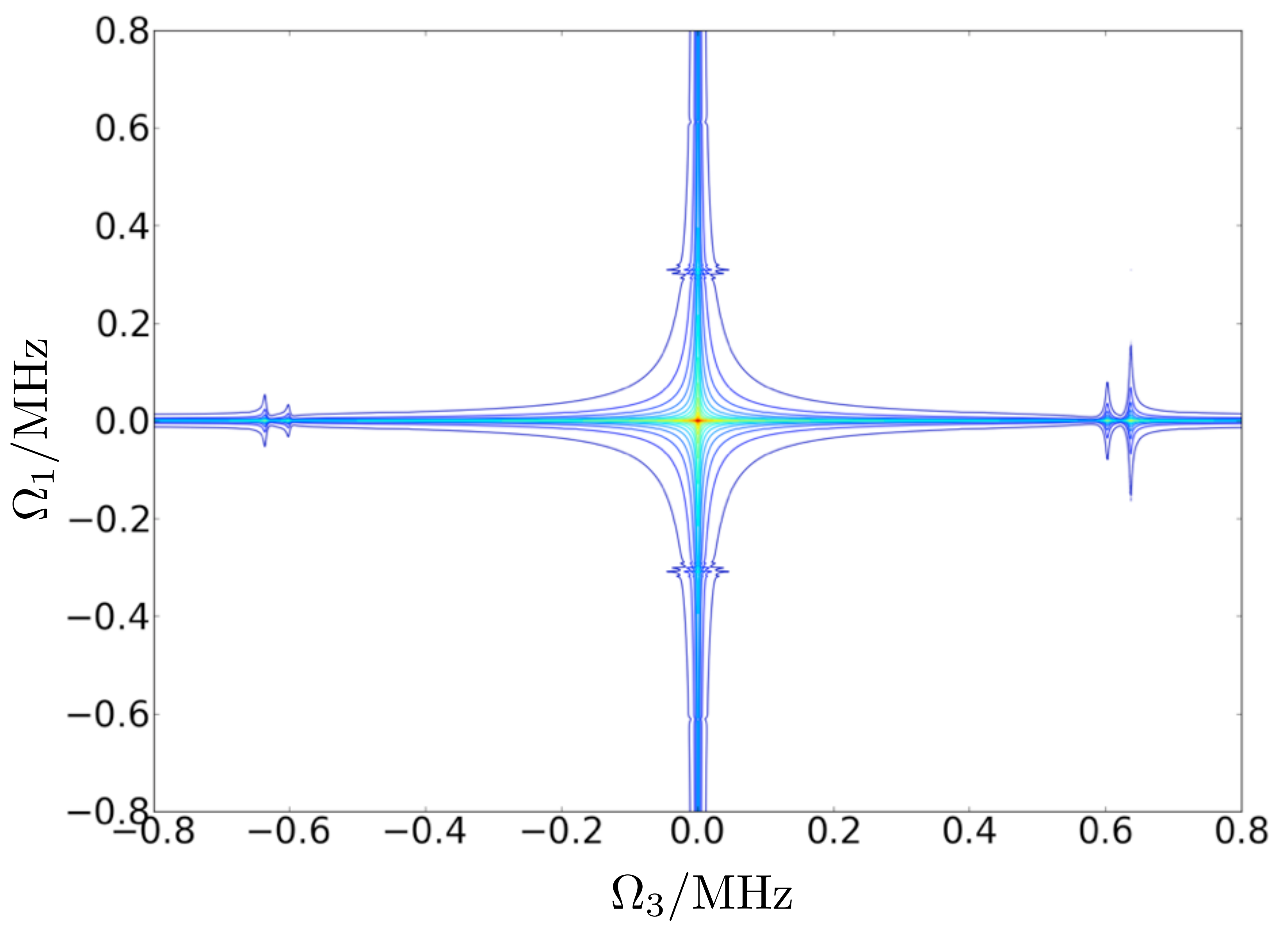}
\caption{Absolute value of the two-dimensional spectrum $|S^{(1, 3)} (\Omega_1, t_2 = 0,\Omega_3)|$, see Eq. (\ref{eq.S^13}), of the pulse sequence~(\ref{eq.signal-definition}) in a two-ion chain, with the Hamiltonian~(\ref{eq.H}) and parameters $\nu_x=1\,$MHz, $\beta=5\,$kHz, $\Delta = - 50\,$kHz, $g = 5\,$kHz.}
\label{fig.f1f3}
\end{figure}

We finally analyze the signal $S^{(1,3)} (\Omega_1, t_2 = 0, \Omega_3)$, Eq.~(\ref{eq.S^13}), for the case of negative detuning 
$\Delta = - 50$~kHz. Here, the two-spin state $|\Psi_2 \rangle$ forms the ground state, and the zero-spin manifold represents the highest-excited states [see Fig.~\ref{fig.levelscheme}b)]. The signal $S^{(1,3)}$ correlates the frequencies of the two free evolution periods
when the system is fully described 
by populations. Consequently, the main resonance can be found at 
zero frequency, with 
a broadening solely due to the finite propagation time. The signal measures the deviations from the $|\Delta / g| \rightarrow \infty$ limit, as the four resonances that can be found at nonzero frequencies all stem from deviations of the ground state $|\Psi_2 \rangle$ from $|\mathrm{atI}\rangle$, and of $|\psi_0^A \rangle$ from $|\mathrm{phSF} \rangle$. The resonances at $\Omega_1 \approx \pm 350$~kHz yield the dominant correction to $|\mathrm{atI}\rangle$, and the resonance pairs at $\Omega_3 \approx \pm 600$~kHz those to $|\mathrm{phSF} \rangle$ in the simulated realization of the polaritonic Hamiltonian.

\section{Summary \& Outlook}
We have proposed a nonlinear measurement scheme to probe the nonequilibrium dynamics of polaritonic quasiparticles in trapped ion chains. Suitably chosen laser pulses can drive the system between an atomic insulator and a superfluid phase, and create 
superposition states of the two phases. On top of the identification of the contributing eigenstate frequencies, the analysis of the two-dimensional lineshapes in the measurement setup allows for a clear identification and characterization of the states, since the dominant noise source is sensitive to the number of spin excitations present in the system. Hence, the broadening of the resonance along specific axes indicates the discrepancy of electronic excitations between the two coherently superposed states. The variation of the third time delay, which was kept fixed in our present study, would further allow to monitor transport and relaxation processes within the $N$-excitation submanifold.

The measurement scheme proposed in this work does not require individual laser access to the ions and is therefore easily 
scalable towards larger $N$. Local information on the excitations can still be achieved by spatially resolved collection of the fluorescence light, which is standard practice in ion trap experiments. The distinct couplings of the ions to different 
collective modes are then reflected in the locally collected measurement signals.

The system we studied -- a collection of two-level atoms that couple to harmonic motional modes -- could also be interesting for experimental studies of the role of the vibrational backbone for exciton transport, which is believed to represent an important mechanism for biomolecular transport processes \cite{Engel07,ReviewPS}. To this end, one would model the direct exciton couplings, which is lacking in the 
polariton model, by adding effective spin-spin interactions \cite{PorrasCiracSpins}. In the context of biomolecules, an electron-phonon coupling that preserves the number of electronic excitations, i.e., a polaronic coupling, would be more appropriate \cite{ReviewPS} than the polaritonic coupling considered in the present study.

\section*{Acknowledments}
M.~G. and F.~S. thank the German National Academic Foundation for support.

\appendix
\section{Derivation of the interaction operator}
\label{appendixA}
In this appendix, we derive Eq.~(\ref{eq.H_I}). Our derivation closely follows standard approaches in the trapped-ion literature, which can be found, for instance in refs. \cite{Leibfried,Retzger}. Its starting point is the Hamiltonian
\begin{align}
H_0 (t) &=  \sum_{k=1}^N \omega_k a_k^{\dagger} a_k + \sum_{\substack{k,l=1\\k < l}}^N t_{kl} \left( a^{\dagger}_k a_l +  a^{\dagger}_l a_k \right) + \omega \sum_{k=1}^N \sigma_k^{+} \sigma_k^-  \notag \\
&\quad+ \frac{\Omega}{2}\sum_{k = 1}^N \left(\sigma_k^+ +\sigma_k^- \right)\left(e^{i( k_L x_k - \omega_L t+\phi)} +  e^{- i( k_L x_k - \omega_L t+\phi)}\right),
\end{align}
which describes the interaction of the ion chain with the external laser field of frequency $\omega_L$, wavevector $\vec{k}_L=k_L\vec{e}_x$ which points along the $x$-axis, and phase $\phi$. We assume spatially homogeneous illumination of all ions, whose optical resonance frequency is given by $\omega$. Note that due to the large separation of the ions (see section \ref{sec.introions}), the spatial dependence of the laser field needs to be taken into account.
We first diagonalize the vibrational Hamiltonian, 
\begin{align}
\sum_{k=1}^N \omega_k a_k^{\dagger} a_k + \sum_{\substack{k,l=1\\k < l}}^N t_{kl} \left( a^{\dagger}_k a_l +  a^{\dagger}_l a_k \right) &= \sum_{n = 1}^N \nu_n b^{\dagger}_n b_n,
\end{align}
and express the motional dependence as a function of the Lamb-Dicke parameters $\eta_n = k_L\sqrt{1/2m\nu_n}$ as \cite{Retzger}
\begin{align}
\exp \left(i k_L x_k \right)= \exp \left( i  \sum_n \eta_n(c_{kn} b_n +c^{\ast}_{kn}  b^{\dagger}_n) \right),
\end{align}
where the spatial orientation of the laser is adjusted such that only the vibrational modes along one direction are addressed \cite{Leibfried}.

The excitation frequency is set on resonance with the red sideband corresponding to the vibrational eigenmode $\nu_m$, i.e., $\omega_L = \omega - \nu_m$. Next we employ a rotating wave approximation and change to the interaction picture, employing the transformation
\begin{align}
U_1 (t) &= \exp\left(- i\left[\omega \sum_k \sigma_k^+ \sigma_k^- + \sum_n \nu_n b^{\dagger}_n b_n\right] t\right),
\end{align}
resulting in
\begin{align}
H_1 (t) &=\frac{\Omega}{2} \sum_k \sigma_k^+ e^{i \nu_m t}e^{-i\phi} \exp\left( i  \sum_n \eta_n(c_{kn} b_n e^{-i\nu_n t}+c^{\ast}_{kn}  b^{\dagger}_ne^{i\nu_n t}) \right) + \mathrm{H. c.}\label{eq.H_1t}
\end{align}
We shall consider the Lamb-Dicke limit \cite{Leibfried}, $\eta_n \sqrt{\langle (b_n+b^{\dagger}_n)^2\rangle} \ll 1$, which is reached when the amplitude of the ion's motion is much less than $1/k$. In this case we may approximate the exponential in Eq.~(\ref{eq.H_1t}) by its Taylor series to linear order,
\begin{align}
\exp\left( i  \sum_n \eta_n(c_{kn} b_n e^{-i\nu_n t}+c^{\ast}_{kn}  b^{\dagger}_ne^{i\nu_n t}) \right) &\approx 1+ i  \sum_n \eta_n(c_{kn} b_n e^{-i\nu_n t}+c^{\ast}_{kn}  b^{\dagger}_ne^{i\nu_n t}).
\end{align}
Retaining only non-oscillatory terms in Eq.~(\ref{eq.H_1t}), we then arrive at
\begin{align}
H_1 (t) &\simeq i \frac{\eta_m \Omega}{2}e^{-i\phi}\left( \sum_{k = 1}^N c_{k m} \sigma_k^+ \right) b_m + \mathrm{H.c.}
\end{align}
For $m=1$ we finally obtain Eq.~(\ref{eq.H_I}) by identifying $J_+= \sum_{k = 1}^N c_{k 1} \sigma_k^+$ and $\widetilde{\Omega}=\eta_1\Omega$.

\subsection{The case $N = 2$}
In the present of case of a two-ion chain, we have two delocalized eigenmodes, which are given by the operators
\begin{align}
b_2 &= \frac{1}{\sqrt{2}} \left( a_1 + a_2 \right),
\end{align}
the center-of-mass (high-energy) mode, and
\begin{align}
b_1 &= \frac{1}{\sqrt{2}} \left( a_1 - a_2 \right),
\end{align}
the breathing (low-energy) mode. Hence, for $N=2$ the coefficients in Eq.~(\ref{eq.H_I}) are $c_{11} = - c_{21} = 1 / \sqrt{2}$.

\end{document}